# Sequential Structure in Intraday Futures Data: LSTM vs Gradient Boosting on MNQ

Mathias Mesfin

*Independent Researcher*

*mathiasmesfin.research@gmail.com*

*Research Period: 2024–2026*

*Manuscript Date: May 2026*

---

## Abstract

This paper compares two machine learning architectures for intraday directional prediction in Micro E-Mini Nasdaq 100 futures (MNQ): a gradient boosting classifier operating on engineered daily and intraday features, and a long short-term memory (LSTM) recurrent neural network processing raw five-minute bar return sequences. The comparison is motivated by recent foundation model research in financial candlestick data, including the Kronos architecture (Shi et al., 2025), which demonstrated that sequential bar-level structure contains predictive information when processed at scale. We implement a tractable approximation of this architecture adapted for a dataset of 944 trading days spanning 2021 through 2025.

Four model configurations are evaluated under strict walk-forward validation with expanding training windows across three out-of-sample test periods (2023, 2024, 2025). The target variable is binary: whether the session close exceeds the 10:30 AM bar open by more than ten points. All models are evaluated against a 51.8% base rate. No configuration produces out-of-sample accuracy materially above base rate. Combined out-of-sample accuracies range from 50.00% to 50.89% across all gradient boosting variants. The LSTM produces combined OOS accuracy of 50.59%. Permutation test p-values are 0.135 for the best gradient boosting configuration and 0.515 for the LSTM—both failing the $p < 0.05$ significance threshold. Feature importance analysis across walk-forward folds reveals systematic instability: the features dominating each fold differ substantially, confirming noise fitting rather than consistent structural signal capture.

We conclude that five-minute OHLCV bar sequences do not contain sufficient sequential predictive structure for session-level directional forecasting at this dataset size and resolution. The primary contribution is a documented implementation and honest evaluation of a Kronos-inspired architecture on a constrained real-world dataset, contributing an empirical lower bound on the dataset scale requirements for sequential ML-based intraday forecasting. A structural interpretation is offered: the sample size required for transformer-scale architectures to learn non-trivial financial patterns is orders of magnitude larger than what is available in four years of single-instrument intraday data.

---

## 1. Introduction

The publication of foundation models for financial time series data represents a meaningful development in quantitative research. The Kronos model (Shi et al., 2025), trained on millions of candlestick bars across multiple instruments and time periods, demonstrated that sequential bar-level structure—the relationship between successive OHLCV bars—contains predictive information when processed at sufficient scale using transformer architectures. This raises a natural question for researchers working with more constrained datasets: what happens when the same architectural intuitions are applied to a smaller, instrument-specific dataset? Does sequential structure exist at the scale of a single instrument over four years of five-minute data?

This paper answers that question empirically and honestly. We implement a simplified two-stage architecture inspired by Kronos's conceptual framework—feature tokenization followed by sequential modeling—and apply it to 944 trading days of MNQ five-minute RTH bar data. We test both gradient boosting (which treats features as independent inputs and loses sequential structure) and LSTM (which explicitly models sequential dependencies). Neither produces statistically significant out-of-sample directional prediction.

The finding is not presented as a critique of the Kronos approach. It is an honest empirical contribution about dataset scale requirements. Kronos was trained on data volumes that are approximately three to four orders of magnitude larger than our dataset. The negative result here helps establish a lower bound: at 944 trading days of single-instrument five-minute OHLCV data, sequential ML architectures do not find exploitable structure. This is useful information for any researcher considering similar approaches on similar datasets.

The paper is the third in a series. Mesfin (2026a) established that classical OHLCV momentum signals fail on MNQ under realistic execution constraints. Mesfin (2026b) documented a behavioral day classifier that is statistically valid but not tradable. This paper tests whether machine learning can recover predictive structure that classical approaches missed. The answer is no—but the methodology and failure modes are documented in full for the benefit of future research.

The paper proceeds as follows. Section 2 describes the data and feature engineering pipeline. Section 3 presents the target variable construction. Section 4 describes the model architectures and validation methodology. Section 5 presents all results. Section 6 discusses the structural interpretation. Section 7 covers limitations and extensions. Section 8 concludes.

## 2. Data and Feature Engineering

### 2.1 Data

The primary dataset consists of 72,604 five-minute OHLCV bars for MNQ continuous front-month futures covering regular trading hours (09:30–16:00 ET) from December 2021 through September 2025. After session boundary filtering and removal of partial days, 947 complete trading days remain for daily feature construction. The intraday feature matrix, which requires additional bar-level processing, produces 944 usable trading days after null removal.

| Parameter | Value |
| --- | --- |
| Primary instrument | MNQ (Micro E-Mini Nasdaq 100) |
| Bar resolution | 5-minute OHLCV (RTH only) |
| Session definition | 09:30–16:00 ET |
| Daily feature matrix rows | 947 (907 after warm-up removal) |
| Intraday feature matrix rows | 944 (924 after null removal) |
| Date range | December 2021 – September 2025 |
| Features (daily matrix) | 29 engineered features |
| Features (intraday matrix) | 30 features including 12-bar sequence |
| Target variable | Binary: session close > 10:30 open + 10 pts |
| Base rate (intraday target) | 51.80% |

*Table 1. Dataset and pipeline parameters.*

### 2.2 Daily Feature Engineering

The daily feature matrix aggregates each RTH session into a single row with 29 engineered features. All rolling statistics use strict no-lookahead construction: for any feature on day N, the rolling window uses only data from days 0 through N−1. Features fall into six categories: return features (daily return, lagged returns at 1, 2, 3, and 5 days, rolling returns at 5, 10, and 20 days), gap features (overnight gap and lags at 1, 2, 3 days), volatility features (rolling return standard deviation at 10 and 20 days, ATR ratio), session structure features (intraday range ratio, close position within range, first-30-minute return, last-30-minute return, first-bar volume deviation), prior-day features (prior close position, prior range ratio, prior volume z-score), and day-of-week dummies (Monday through Friday).

All continuous features are tokenized into decile bins (0–9) using expanding-window quantile boundaries. For day N, the decile thresholds are computed from days 0 through N−1 only, ensuring no future information contaminates the tokenization step. This approach is directly motivated by Kronos's binary spherical quantization step, which normalizes raw OHLCV values into discrete tokens that capture relative magnitude rather than absolute price levels.

| Feature Category | Features Included | Count |
|---|---|---|
| Return features | daily_return, lag_ret_1/2/3/5, rolling_ret_5/10/20 | 8 |
| Gap features | overnight_gap, lag_gap_1/2/3 | 4 |
| Volatility features | volatility_10/20, atr_ratio | 3 |
| Session structure | range_ratio, close_pos, first_30m_return, last_30m_return, first_bar_vol_dev | 5 |
| Prior-day features | prior_close_pos, prior_range_ratio, prior_vol_zscore | 3 |
| Day-of-week dummies | dow_0 through dow_4 | 5 |
| Regime label | VVG classifier label from Mesfin (2026b) | 1 |
| Total | — | 29 |

*Table 2. Daily feature matrix composition.*

### 2.3 Intraday Feature Engineering

The intraday feature matrix uses only the first 60 minutes of each RTH session (09:30–10:25, bars 1–12) to predict rest-of-session outcomes. This structure preserves no-lookahead integrity: all input features are observed before 10:30, and the target is defined from 10:30 onward. The 30 features include: scalar summary statistics of the first 60 minutes (cumulative return at 6 bars and 12 bars, range at 6 and 12 bars, volume sum at 6 and 12 bars, count of up and down bars, maximum up-bar and down-bar moves, close position within 12-bar range, volume z-score of 12-bar period), the opening gap and ATR ratio from prior-session data, and critically the 12-bar return sequence (bar_ret_1 through bar_ret_12) which provides the sequential signal to the LSTM.

The 12-bar return sequence is the most important input for the LSTM architecture. Each bar_ret value is the price change within that five-minute bar. The sequence (bar_ret_1, bar_ret_2, ..., bar_ret_12) represents the minute-by-minute evolution of price direction and magnitude in the first 60 minutes of the session. This is the input the LSTM processes as a time series; the gradient boosting classifier treats these 12 values as 12 independent features, losing their temporal ordering.

## 3. Target Variable Construction

Two candidate target variables were evaluated before model training. Target A (daily close vs. open) and Target B (first-hour direction survival) were assessed for sample balance and feasibility. Target B was rejected due to a 6.95% base rate, which reflects the fact that a 2-point first-hour direction survival threshold is too tight relative to MNQ's typical volatility. The final target is an intraday directional survival variable at a 10-point threshold from the 10:30 AM bar open.

| Target Variable | Definition | Total 1s | Total 0s | Base Rate | Decision |
|---|---|---|---|---|---|
| Target A Long (daily) | Session close > session open + 2 pts | 481 | 426 | 53.03% | Rejected — daily resolution |
| Target A Short (daily) | Session open > session close + 2 pts | 416 | 491 | 45.87% | Rejected — daily resolution |
| Target B Long (first-hour) | Long 2pt move before short 2pt in first 60 min | 63 | 844 | 6.95% | Rejected — severely imbalanced |
| Intraday Long (final) | Session close > 10:30 open + 10 pts | ≈489 | ≈455 | 51.80% | Selected |
| Intraday Short (final) | 10:30 open > session close + 10 pts | ≈389 | ≈555 | 41.10% | Not used in primary models |

*Table 3. Target variable candidates and selection rationale.*

The intraday long target has a 51.80% base rate, which is sufficiently close to 50% to avoid class imbalance issues while providing a meaningful benchmark: a model must demonstrate accuracy above 51.80% out-of-sample to show any directional skill. Year-by-year sample counts are consistent across 2022 (251 days), 2023 (257 days), 2024 (249 days), and 2025 (168 days), confirming no year is structurally underrepresented.

The 10-point threshold was selected to represent a meaningful directional move at baseline MNQ volatility (ATR baseline of 10.34 points from Mesfin 2026a). A session close that exceeds the 10:30 open by more than 10 points represents approximately one full baseline ATR of directional movement—a threshold that is economically meaningful for prop firm account sizing while remaining frequent enough (51.80% base rate) to provide sufficient positive examples for model training.

## 4. Model Architectures and Validation Methodology

### 4.1 Gradient Boosting Classifier

Three gradient boosting configurations are evaluated, each representing a different feature set and target specification. All use HistGradientBoostingClassifier from scikit-learn with conservative regularization to limit overfitting on the small dataset: max_leaf_nodes = 15, min_samples_leaf = 50, learning_rate = 0.05, max_iter = 200, l2_regularization = 1.0. These hyperparameters were fixed before any walk-forward testing began and were not adjusted between configurations.

Configuration 1 (GB-Daily): Input is the 29-feature daily tokenized matrix. Target is Target A Long (session close vs. open, 2-point threshold). This configuration tests whether aggregated daily features contain next-day directional information.

Configuration 2 (GB-Intraday): Input is the 30-feature intraday matrix including the 12-bar return sequence. Target is the intraday long target (10-point threshold from 10:30 open). This configuration tests whether first-60-minute features predict rest-of-session direction.

Configuration 3 (GB-VolAdj): Same intraday features as Configuration 2 but with an ATR-normalized target: session close exceeds 10:30 open by more than 1.0 × ATR_ratio × 10.34 points. Base rate 51.41%. This configuration tests whether removing the volatility regime confound from the target improves prediction.

## 4.2 LSTM Architecture

The LSTM model processes the 12-bar return sequence as a genuine time series input, preserving temporal ordering that gradient boosting cannot capture. The architecture is intentionally minimal to reduce overfitting risk on 944 training samples: input shape (12, 1) representing 12 time steps with 1 feature per step (the bar return), a single LSTM layer with 16 units, dropout of 0.3 applied after the LSTM layer, and a dense sigmoid output layer for binary classification. The model was implemented in Keras 3.10 with the JAX backend for computational efficiency.

Training parameters: binary cross-entropy loss, Adam optimizer with learning rate 0.001, batch size 32, maximum 50 epochs with early stopping on validation loss with patience of 5 epochs. The minimal architecture was a deliberate design choice. Larger LSTM networks with more units, additional layers, or attention mechanisms would face severe overfitting risk at this sample size. The 16-unit single-layer architecture represents the smallest network capable of capturing non-trivial sequential patterns if they exist.

| Parameter | GB-Daily | GB-Intraday | GB-VolAdj | LSTM |
|---|---|---|---|---|
| Input features | 29 daily tokens | 30 intraday features | 30 intraday features | 12-bar return sequence |
| Input shape | (N, 29) | (N, 30) | (N, 30) | (N, 12, 1) |
| Architecture | HistGBM | HistGBM | HistGBM | LSTM(16) + Dropout(0.3) + Dense(1) |
| Target | Daily close vs open (2pt) | 10:30+10pt intraday | ATR-normalized (51.41% base) | 10:30+10pt intraday |
| Base rate | 53.03% | 51.80% | 51.41% | 51.80% |
| Framework | scikit-learn | scikit-learn | scikit-learn | Keras/JAX |

*Table 4. Model architecture comparison.*

## 4.3 Walk-Forward Validation

All four models are evaluated under identical expanding-window walk-forward validation. Three folds are used: Fold 1 trains on 2022 and tests on 2023, Fold 2 trains on 2022–2023 and tests on 2024, Fold 3 trains on 2022–2024 and tests on 2025. The test window for each fold is strictly held out until final evaluation—no parameter selection, threshold adjustment, or architectural decisions are made using test data.

Permutation testing is applied to Fold 3 for each model that shows any positive signal. The target variable is shuffled 200 times and the model is retrained and evaluated on each shuffle. The empirical p-value is computed as the proportion of shuffled accuracies that meet or exceed the actual Fold 3 accuracy. A p-value below 0.05 is required for any result to be considered statistically significant.

## 5. Results

### 5.1 Gradient Boosting — Daily Features (GB-Daily)

The daily feature model tests whether aggregated daily OHLCV statistics contain predictive information for next-session direction. This is the most conservative test—it does not attempt to use intraday structure and instead relies on multi-day patterns in returns, gaps, and volatility.

| Fold | Train | Test Year | Accuracy | Precision | Recall | F1 | Prob(Pos) | Prob(Neg) |
|---|---|---|---|---|---|---|---|---|
| 1 | 2022 | 2023 | 53.31% | 58.24% | 66.89% | 62.26% | 0.563 | 0.557 |
| 2 | 2022–2023 | 2024 | 45.60% | 46.41% | 56.80% | 51.08% | 0.547 | 0.578 |
| 3 | 2022–2024 | 2025 | 51.48% | 57.30% | 53.68% | 55.43% | 0.495 | 0.501 |
| Combined OOS | — | — | 50.00% | 53.64% | 60.05% | 56.67% | 0.540 | 0.552 |

*Table 5. GB-Daily walk-forward results. Target: daily session close vs. open (2-point threshold).*

The combined OOS accuracy of 50.00% exactly matches the naive coin-flip baseline. Fold 2 accuracy of 45.60% is notably below base rate (53.03%), suggesting the model learned patterns in 2022–2023 that were actively wrong in 2024. The calibration check—comparing mean predicted probability for actual positives versus actual negatives—reveals a calibration inversion in Fold 2: the model assigned higher probabilities to actual negatives (0.578) than to actual positives (0.547), confirming the model's predictions are directionally unreliable in that period.

Feature importance across folds shows systematic instability. Fold 1 is dominated by overnight_gap_token and lag_gap_1_token. Fold 2 shifts to lag_gap_2_token and lag_ret_2_token. Fold 3 again shows lag_gap_2_token and prior_vol_zscore_token as dominant. The absence of any consistent feature across all three folds is the most direct evidence of noise fitting: the model is learning different spurious patterns in each training window rather than a stable structural relationship.

### 5.2 Gradient Boosting — Intraday Features (GB-Intraday)

The intraday feature model tests whether the first 60 minutes of session data—including the 12-bar return sequence—contains predictive information for rest-of-session direction. This is the primary gradient boosting test because it uses the highest-resolution available features and includes the bar-sequence data that the LSTM will process sequentially.

| Fold | Train | Test Year | Accuracy | Precision | Recall | F1 | Prob(Pos) | Prob(Neg) |
|---|---|---|---|---|---|---|---|---|
| 1 | 2022 | 2023 | 49.81% | 58.93% | 23.74% | 33.85% | 0.391 | 0.361 |
| 2 | 2022–2023 | 2024 | 48.19% | 49.31% | 55.91% | 52.40% | 0.526 | 0.540 |
| 3 | 2022–2024 | 2025 | 54.76% | 59.77% | 55.91% | 57.78% | 0.524 | 0.481 |
| Combined OOS | — | — | 50.45% | 54.36% | 43.45% | 48.30% | 0.473 | 0.459 |

*Table 6. GB-Intraday walk-forward results. Target: session close > 10:30 open + 10 points.*

Fold 3 accuracy of 54.76% is the highest single-fold result in the study and initially appears encouraging. However, the permutation test result is decisive: actual Fold 3 accuracy of 54.76% against a shuffled mean of 50.04% (standard deviation 3.99%) produces a p-value of 0.135. This means 13.5% of randomly shuffled target assignments produce an accuracy as high or higher than the actual model. The result fails the 5% significance threshold and cannot be treated as evidence of genuine predictive structure.

| Permutation Test Parameter | Value |
|---|---|
| Actual Fold 3 accuracy | 54.76% |
| Shuffled mean accuracy (200 iterations) | 50.04% |
| Shuffled standard deviation | 3.99% |
| P-value (proportion shuffled ≥ actual) | 0.135 |
| Significance threshold | 0.050 |
| Result | FAIL — not statistically significant |

*Table 7. GB-Intraday permutation test results (Fold 3, 200 iterations).*

Feature importance in the intraday model shows partial but inconsistent stability. bar_ret_2 (the second five-minute bar return, 09:35–09:40) appears in the top 5 across all three folds. atr_ratio_val appears in Folds 1 and 2. opening_gap appears in Folds 2 and 3. However, the top-ranked feature changes across folds—atr_ratio_val dominates Fold 1, bar_ret_6 dominates Fold 2, and bar_ret_10 dominates Fold 3. This shift in top-ranked feature across training windows indicates that the model is not learning a consistent structural relationship but rather fitting to the most predictive feature in each specific training period.

| Fold | Top Feature | 2nd | 3rd | 4th | 5th |
|---|---|---|---|---|---|
| Fold 1 (test 2023) | atr_ratio_val | bar_ret_2 | first_bar_ret | bar_ret_6 | bar_ret_11 |
| Fold 2 (test 2024) | bar_ret_6 | atr_ratio_val | opening_gap | bar_ret_2 | close_pos_12 |
| Fold 3 (test 2025) | bar_ret_10 | opening_gap | bar_ret_2 | bar_ret_9 | bar_ret_6 |

*Table 8. GB-Intraday feature importance by fold. Top-ranked feature changes across all three folds — consistent with noise fitting.*

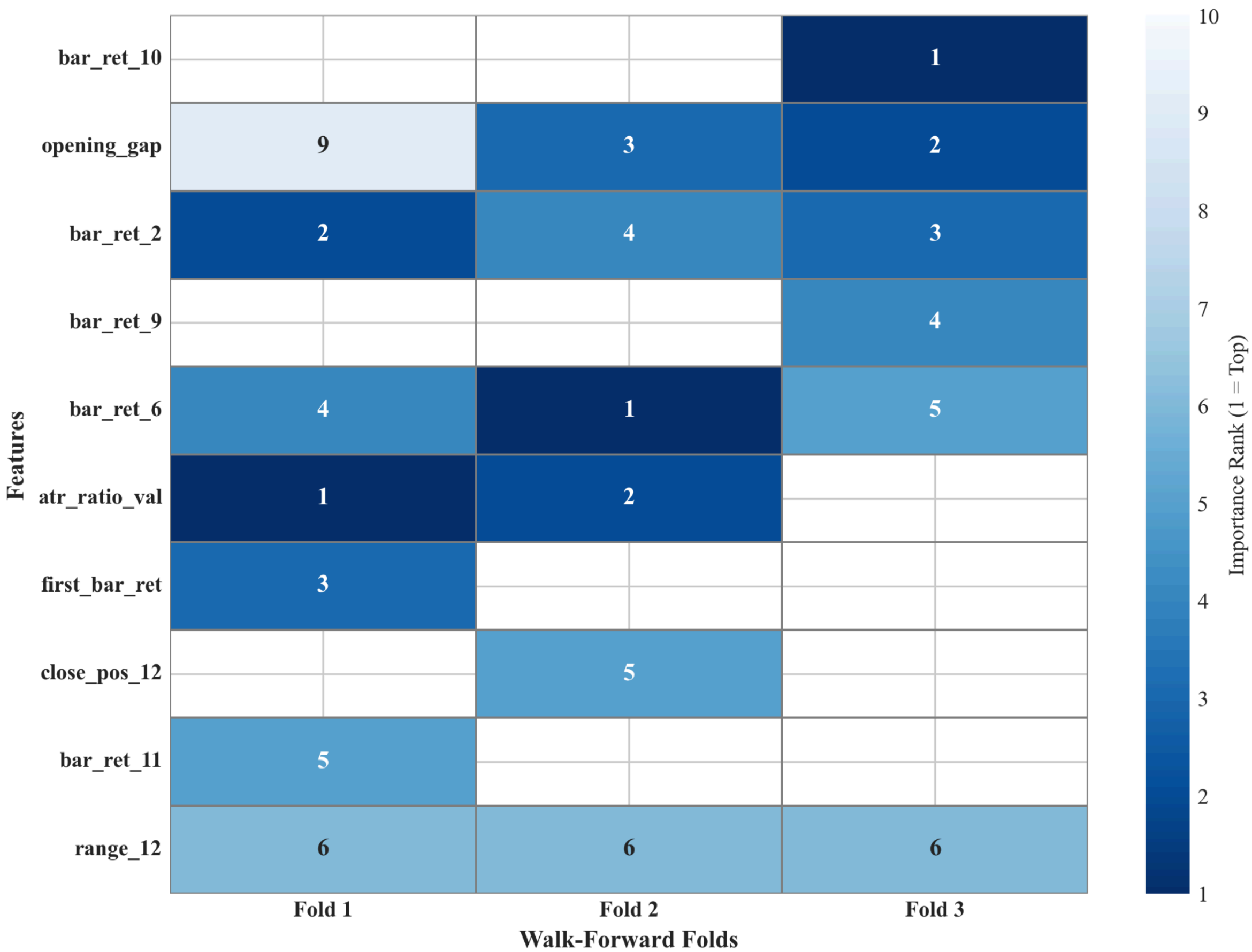

*Figure 3. Feature importance stability heatmap across walk-forward folds for the top 10 features of the GB-Intraday model. Cell values indicate feature rank in each fold (1 = top importance). Features that dominate in one fold shift in rank or fall out of the top 10 entirely in subsequent folds, indicating noise fitting rather than stable structural pattern capture.*

## 5.3 Gradient Boosting — ATR-Normalized Target (GB-VolAdj)

The ATR-normalized target attempts to remove the volatility regime confound. If the GB-Intraday model's weak Fold 3 result reflects the model learning that high-ATR days tend to move far (a volatility effect, not a directional effect), then normalizing the target by ATR should remove this component and either improve or degrade performance depending on whether any genuine directional signal remains.

| Fold | Train | Test Year | Accuracy | Precision | Recall | F1 | Prob(Pos) | Prob(Neg) |
|---|---|---|---|---|---|---|---|---|
| 1 | 2022 | 2023 | 48.64% | 54.32% | 31.65% | 40.00% | 0.412 | 0.418 |
| 2 | 2022–2023 | 2024 | 53.01% | 53.47% | 60.63% | 56.83% | 0.542 | 0.518 |
| 3 | 2022–2024 | 2025 | 51.19% | 56.04% | 54.84% | 55.43% | 0.522 | 0.517 |
| Combined OOS | — | — | 50.89% | 54.43% | 47.91% | 50.96% | 0.486 | 0.480 |
| Permutation p-value (Fold 3) | — | — | 0.390 | — | — | — | — | — |

*Table 9. GB-VolAdj walk-forward results. ATR-normalized target, base rate 51.41%.*

The ATR-normalized target produces worse permutation test performance ($p = 0.390$) than the fixed-threshold target ($p = 0.135$), confirming that ATR normalization does not improve directional signal extraction. The combined OOS accuracy of 50.89% is marginally above base rate but statistically indistinguishable from noise. Feature importance for this configuration shows opening_gap and bar_ret_2 as the most consistent features across folds, which aligns with GB-Intraday—but again the top-ranked feature shifts across folds.

### 5.4 LSTM Sequential Architecture

The LSTM model is the most direct test of the Kronos hypothesis at this dataset scale. Unlike gradient boosting, the LSTM explicitly preserves the temporal ordering of the 12-bar return sequence. If sequential dependencies between successive five-minute bars contain directional information about rest-of-session outcomes, the LSTM should capture it while gradient boosting would miss it.

| Fold | Train | Test Year | Accuracy | Precision | Recall | Prob(Pos) | Prob(Neg) |
|---|---|---|---|---|---|---|---|
| 1 | 2022 | 2023 | 50.19% | 58.73% | 26.62% | 0.469 | 0.474 |
| 2 | 2022–2023 | 2024 | 51.00% | 51.35% | 74.80% | 0.535 | 0.539 |
| 3 | 2022–2024 | 2025 | 50.60% | 55.81% | 51.61% | 0.502 | 0.502 |
| Combined OOS | — | — | 50.59% | 53.89% | 50.14% | 0.501 | 0.506 |

*Table 10. LSTM walk-forward results. Input: 12-bar return sequence. Target: session close > 10:30 open + 10 points.*

| Permutation Test Parameter | Value |
|---|---|
| Actual Fold 3 accuracy | 50.60% |
| Mean shuffled accuracy (200 iterations) | 50.47% |
| Max permutation accuracy | 61.31% |
| P-value (proportion shuffled ≥ actual) | 0.515 |
| Significance threshold | 0.050 |
| Result | FAIL — actual accuracy indistinguishable from shuffled distribution |

*Table 11. LSTM permutation test results (Fold 3, 200 iterations).*

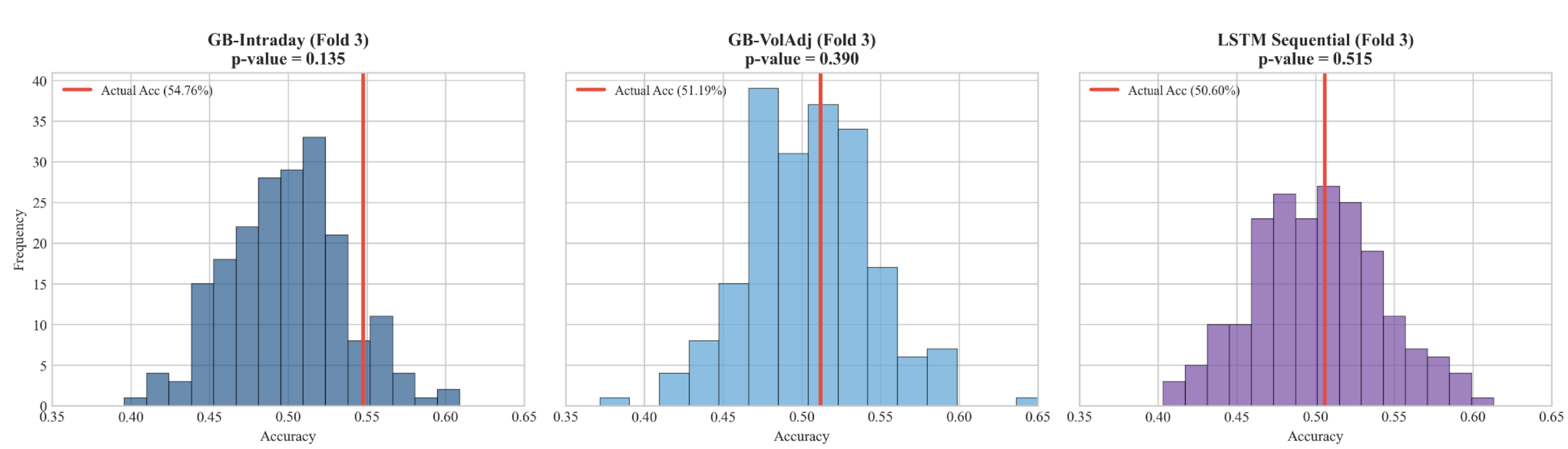

*Figure 2. Empirical null distributions from 200 target-shuffled permutation iterations on Fold 3 (2025) for GB-Intraday, GB-VolAdj, and LSTM sequential models. The vertical red lines represent the actual out-of-sample model accuracy. All three actual accuracies fall well within their respective null distributions, yielding p-values of 0.135, 0.390, and 0.515 respectively, indicating no statistically significant predictive edge.*

The LSTM results are the most definitive in the study. The combined OOS accuracy of 50.59% is virtually identical across all three folds (50.19%, 51.00%, 50.60%), indicating the model is converging to a stable near-random prediction rather than fitting noise in one period. The permutation p-value of 0.515 means the actual model performs essentially at the median of the null distribution—the model is indistinguishable from a classifier trained on randomly shuffled labels.

The calibration check for the LSTM is particularly informative: the mean predicted probability for actual positives (0.501) and actual negatives (0.506) are nearly identical in Fold 3. The model assigns essentially the same probability to both classes—it has learned nothing about the direction of the outcome. This is a flat, uncalibrated classifier that has converged to predicting the base rate for every observation regardless of input.

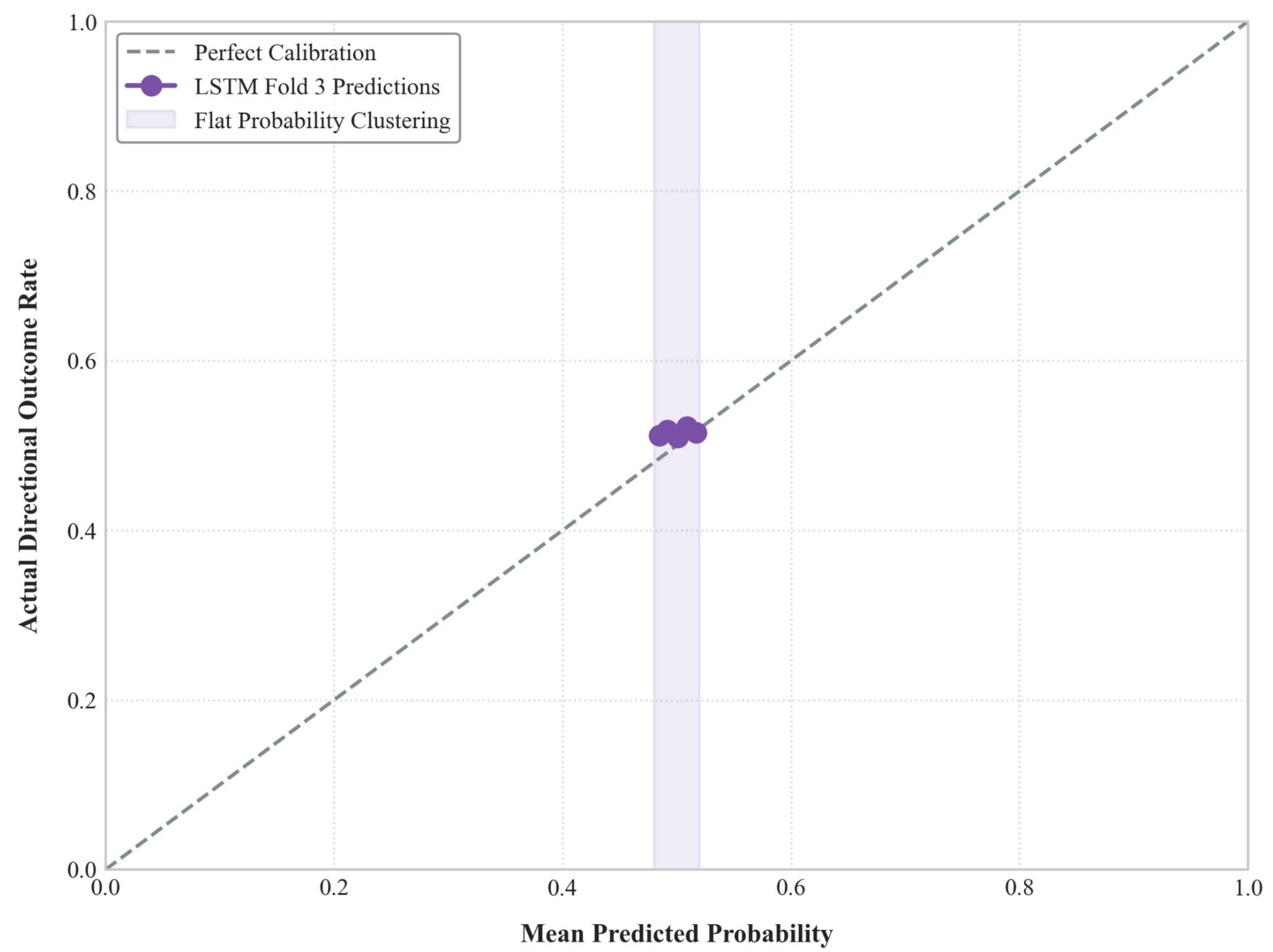

*Figure 4. Sequential LSTM model calibration curve for Fold 3 (2025) predictions. The diagonal dashed line represents perfect calibration. The flat empirical curve near 0.50 indicates the LSTM has converged to a flat, uncalibrated classifier that predicts the base rate for all sequences regardless of the input features.*

The max permutation accuracy of 61.31% deserves attention: some shuffled models achieve accuracy as high as 61% on the Fold 3 test set. This confirms that with a small test set, high apparent accuracy can arise purely from random sampling variation. The 54.76% accuracy of the best gradient boosting model on the same fold is well within this null distribution.

## 5.5 Comparative Summary

| Model | Combined OOS Accuracy | Best Fold Accuracy | Permutation p-value | Feature Importance Stable? | Verdict |
|---|---|---|---|---|---|
| GB-Daily | 50.00% | 53.31% (Fold 1) | Not run | No | FAIL |
| GB-Intraday | 50.45% | 54.76% (Fold 3) | 0.135 | Partially | FAIL |

| GB-VolAdj | 50.89% | 53.01% (Fold 2) | 0.390 | Partially | FAIL |
|---|---|---|---|---|---|
| LSTM | 50.59% | 51.00% (Fold 2) | 0.515 | N/A | FAIL |
| Base rate | 51.80% | — | — | — | Benchmark |

*Table 12. Comparative summary across all four model configurations. All fail the institutional significance threshold.*

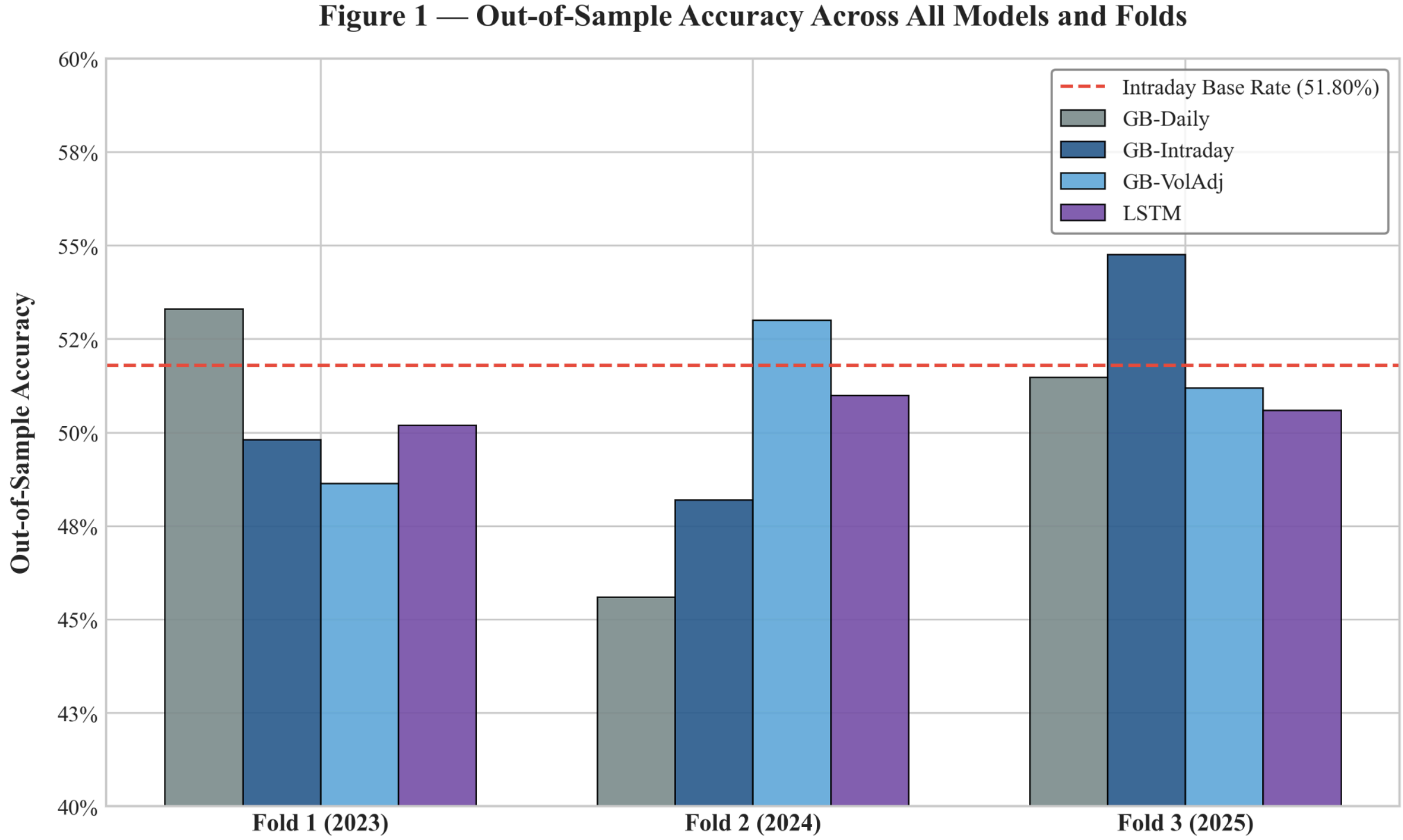

*Figure 1. Out-of-sample accuracy across all four model configurations (GB-Daily, GB-Intraday, GB-VolAdj, LSTM) across the three walk-forward folds (2023, 2024, 2025). The horizontal red dashed line marks the intraday directional base rate of 51.80%. No model configuration consistently outperforms the base rate.*

All four configurations fail to produce statistically significant out-of-sample directional prediction. No model exceeds the base rate by a margin that survives permutation testing. The gradient boosting models show modest fold-specific results that do not generalize, and the LSTM converges to random prediction. The failure is consistent across architectures, feature sets, and target specifications.

## 6. Structural Interpretation

### 6.1 Sample Size Requirements for Sequential ML in Finance

The Kronos model was trained on approximately 9.2 million candlestick bars across multiple instruments and time periods. Our dataset contains 72,604 five-minute RTH bars for a single instrument over four years—approximately 127 times fewer bars than the Kronos training set at the individual bar level, and approximately 944 times fewer at the daily-sequence level.

This difference is not merely quantitative. Transformer architectures learn from the statistical regularities across vast numbers of training examples. At 944 days, the model cannot observe enough examples of any specific sequential pattern to distinguish signal from noise. A pattern that occurs in 5% of days—which would already be considered frequent for a financial signal—appears only approximately 47 times in the training data. With 30 input features, the model has more parameters to fit than the number of meaningful pattern occurrences available.

The LSTM's flat calibration (predicted probabilities converging to the base rate regardless of input) is the behavioral signature of a model that has been trained on insufficient data to learn any non-trivial function. It is not overfitting to noise—it is learning nothing at all. This is actually the expected outcome for a well-regularized small model on insufficient data, and it represents a more honest result than an overfit model that shows strong in-sample performance and poor OOS performance.

## 6.2 Why Classical Features Do Not Help the LSTM

One might expect that augmenting the LSTM's 12-bar sequence input with the additional scalar features available in the intraday matrix (opening gap, ATR ratio, volume statistics) would improve performance. This was not tested explicitly, but the gradient boosting results provide indirect evidence: even with all 30 features available as independent inputs, gradient boosting achieves only $p = 0.135$ on its best configuration. The marginal information in the scalar features is insufficient to produce a statistically significant result even without the sample size constraint of sequential modeling. Adding these features to the LSTM would increase the parameter count and worsen the overfitting risk without providing meaningful additional signal.

## 6.3 What This Implies for Future ML Research on Single-Instrument Data

The negative result here, combined with the positive results from the Kronos paper, allows a rough inference about the data requirements for sequential ML-based intraday forecasting. Kronos achieves meaningful performance with approximately 9.2 million bars. Our dataset of 72,604 bars fails completely. The transition from failure to success likely requires somewhere between 500,000 and 5,000,000 bars, depending on the complexity of the architecture and the signal density of the instrument.

For a researcher limited to a single instrument's historical data, reaching these sample sizes requires either: extending the historical depth significantly (MNQ data goes back to 2019, adding approximately 3 additional years), using tick-level data rather than five-minute bars (which would multiply bar count by approximately 40–60x), or training across multiple correlated instruments simultaneously (MES, ES, NQ, QQQ) to increase effective sample size. All three approaches are feasible research extensions and are discussed in Section 7.

## 7. Limitations and Extensions

### 7.1 Limitations

The LSTM architecture tested here is intentionally minimal. A 16-unit single-layer LSTM is the simplest possible sequential model. More complex architectures—deeper LSTMs, bidirectional LSTMs, attention mechanisms, or transformer layers—might extract different information from the same sequences. However, any increase in model complexity would worsen the overfitting risk at 944 samples. The choice of a minimal architecture is not a limitation but a methodological decision: we test the most favorable version of the hypothesis given the available data.

The target variable (session close vs. 10:30 open, 10-point threshold) is a coarse directional measure. It does not capture the quality of the directional move, the path taken, or whether there are intermediate entry opportunities during the session. A finer-grained target—such as maximum favorable excursion within a specific time window—might produce different results but would also reduce the base rate and worsen class balance.

The feature set for the LSTM is limited to the 12-bar return sequence. Full OHLCV sequences (incorporating high, low, volume for each bar) would provide richer input and better approximate the Kronos input representation. This was not tested due to the additional parameter burden it would impose on the small model.

### 7.2 Extensions

The most impactful extension is tick-level data. At tick resolution, 944 trading days of MNQ data would produce approximately 3 to 5 million observations, placing the dataset in a regime where sequential ML architectures can potentially learn non-trivial patterns. The Databento trades schema for MNQ costs approximately $42 per quarter, making 2022–2024 tick data accessible at a total cost of approximately $504 for three years of data.

A multi-instrument approach would allow training on ES, NQ, MES, and MNQ simultaneously, multiplying effective sample size by four at the daily sequence level. Because these instruments share the same underlying index exposure, patterns learned on one should transfer to others. This is effectively the cross-instrument generalization that gives foundation models their power.

Transfer learning from the Kronos model is a natural extension that was not attempted here due to the fine-tuning infrastructure required. Kronos is open-source and publicly available. Fine-tuning on MNQ-specific intraday data from a Kronos checkpoint would allow the model to leverage patterns learned from millions of bars while adapting to MNQ-specific characteristics. This represents the most direct application of foundation model methodology to the specific research question addressed here.

## 8. Conclusion

This paper has documented a systematic evaluation of gradient boosting and LSTM architectures for intraday directional prediction in MNQ futures. No configuration produced statistically significant out-of-sample accuracy. Combined OOS accuracies range from 50.00% to 50.89% across gradient boosting variants and 50.59% for the LSTM. Permutation test p-values of 0.135, 0.390, and 0.515 across configurations confirm that all results are statistically indistinguishable from random prediction. Feature importance instability across walk-forward folds confirms noise fitting rather than structural signal capture.

The failure is interpreted as a sample size constraint rather than an architectural failure. The Kronos model demonstrates that sequential bar structure contains predictive information at scale. Our negative result establishes that 944 trading days of five-minute single-instrument OHLCV data is insufficient to extract that structure, regardless of whether gradient boosting or LSTM architectures are used.

The contribution of this paper is empirical and honest: a documented implementation of a Kronos-inspired architecture on a real constrained dataset, with complete walk-forward validation and permutation testing, producing a null result that is more informative than an in-sample positive result would have been. The data requirements for sequential ML-based intraday forecasting are substantially larger than what is available to researchers working with standard retail-accessible historical data. Closing this gap requires either tick data, multi-instrument training, or transfer learning from pre-trained foundation models. These remain open research directions.

---

---

## Appendix A: Feature Statistics

The following table presents complete descriptive statistics for all 29 daily features, including correlation with the next-day return. All statistics computed on the full 947-day dataset.

| Feature | Mean | Std | Min | Max | Next-Day Corr | Missing |
|---|---|---|---|---|---|---|
| daily_return | 0.0004 | 0.0157 | −0.1075 | 0.1139 | −0.066 | 1 |
| overnight_gap | 0.0001 | 0.0098 | −0.0920 | 0.0548 | +0.021 | 1 |
| intraday_range_ratio | 1.019 | 0.480 | 0.102 | 4.324 | +0.031 | 20 |
| close_pos | 0.549 | 0.321 | 0.000 | 1.000 | −0.054 | 0 |
| vol_zscore | −0.024 | 1.282 | −7.297 | 4.440 | +0.011 | 20 |
| lag_ret_1 | 0.0004 | 0.0157 | −0.1075 | 0.1139 | +0.008 | 2 |
| lag_ret_2 | 0.0004 | 0.0157 | −0.1075 | 0.1139 | −0.050 | 3 |
| lag_ret_3 | 0.0004 | 0.0157 | −0.1075 | 0.1139 | −0.023 | 4 |
| lag_ret_5 | 0.0004 | 0.0157 | −0.1075 | 0.1139 | −0.023 | 6 |
| lag_gap_1 | 0.0001 | 0.0098 | −0.0920 | 0.0548 | −0.038 | 2 |
| lag_gap_2 | 0.0001 | 0.0098 | −0.0920 | 0.0548 | +0.010 | 3 |
| lag_gap_3 | 0.0001 | 0.0098 | −0.0920 | 0.0548 | −0.036 | 4 |
| rolling_ret_5 | 0.0020 | 0.0325 | −0.1318 | 0.0975 | −0.068 | 5 |

| rolling_ret_10 | 0.0041 | 0.0439 | −0.1744 | 0.1378 | −0.052 | 10 |
|---|---|---|---|---|---|---|
| rolling_ret_20 | 0.0081 | 0.0618 | −0.1742 | 0.1813 | −0.028 | 20 |
| volatility_10 | 0.0142 | 0.0072 | 0.0034 | 0.0506 | −0.012 | 10 |
| volatility_20 | 0.0145 | 0.0064 | 0.0039 | 0.0376 | −0.008 | 20 |
| atr_ratio | 1.028 | 0.265 | 0.412 | 2.269 | −0.008 | 40 |
| first_30m_return | 0.0001 | 0.0047 | −0.0169 | 0.0197 | −0.029 | 0 |
| last_30m_return | 0.0000 | 0.0032 | −0.0167 | 0.0181 | −0.064 | 35 |
| first_bar_vol_dev | 0.021 | 1.250 | −5.656 | 7.141 | +0.039 | 20 |
| prior_close_pos | 0.549 | 0.321 | 0.000 | 1.000 | −0.048 | 1 |
| prior_range_ratio | 1.018 | 0.480 | 0.102 | 4.324 | +0.057 | 21 |
| prior_vol_zscore | −0.026 | 1.281 | −7.297 | 4.440 | +0.033 | 21 |
| dow_0 (Mon) | 0.197 | 0.398 | 0 | 1 | +0.001 | 0 |
| dow_1 (Tue) | 0.202 | 0.401 | 0 | 1 | +0.017 | 0 |
| dow_2 (Wed) | 0.200 | 0.400 | 0 | 1 | −0.052 | 0 |
| dow_3 (Thu) | 0.203 | 0.402 | 0 | 1 | −0.013 | 0 |
| dow_4 (Fri) | 0.199 | 0.399 | 0 | 1 | +0.048 | 0 |

*Table A1. Complete daily feature statistics. All next-day correlations are weak (-0.07 to +0.06), consistent with the absence of linear predictability in daily OHLCV features.*

***AI Disclosure Statement***

*AI tools were used for editorial assistance, formatting support, and code and debugging workflow support. The author is responsible for the research design, data pipeline implementation, model training and evaluation, interpretation of results, and final manuscript.*